# A shark in the stars: astronomy and culture in the Torres Strait

Duane Hamacher
*Nura Gili Indigenous Programs Unit, University of New South Wales, Sydney.*

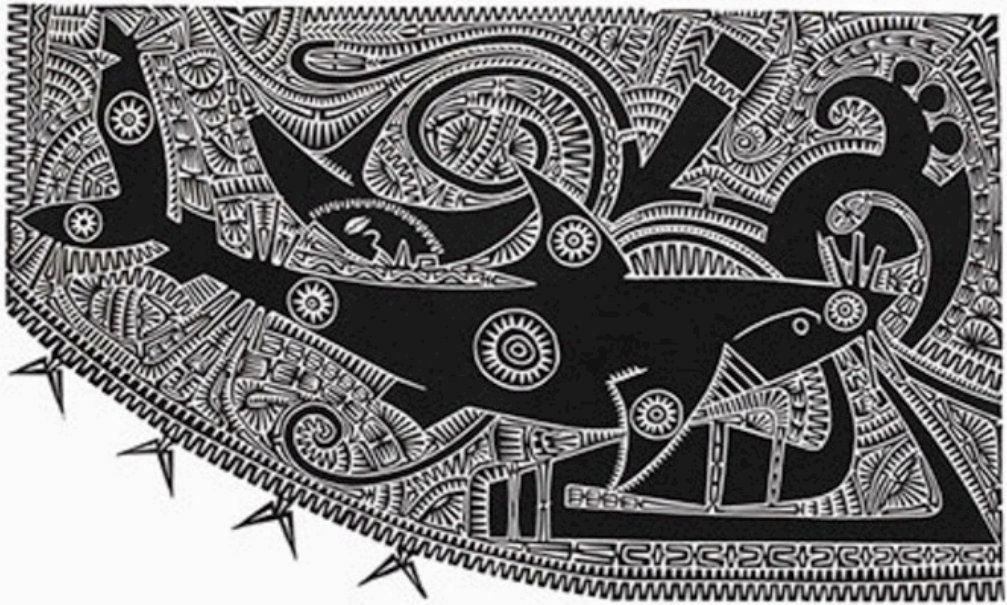

*Torres Strait Islanders use constellations, such as the shark 'Baidam' pictured here, for practical purposes. Credit: Brian Robinson.*

Technology has, without doubt, expanded our understanding of space. The Voyager 1 space probe is on the brink of leaving our solar system. Massive telescopes have discovered blasts of fast radio bursts from 10 billion light years away. And after a decade on Mars, a Rover recently found evidence for an early ocean on the Red Planet.

But with every new advance, it's also important to remember the science of astronomy has existed for thousands of years and forms a vital part of Indigenous Australian culture, even today. As an example, let's explore the astronomy of the Torres Strait Islanders, an Indigenous Australian people living between the tip of Cape York and Papua New Guinea.

Torres Strait Islanders are a Melanesian sea-faring people whose traditional country comprises 48,000 square kilometres of shallow waters and more than 250 islands of differing geological formation, of which 14 are inhabited.

Culturally, the islands are divided into five groups, represented by the five-pointed star on the Torres Strait flag:

- Top Western Islands
- Western Islands
- Lower Western Islands
- Central Islands
- Eastern Islands

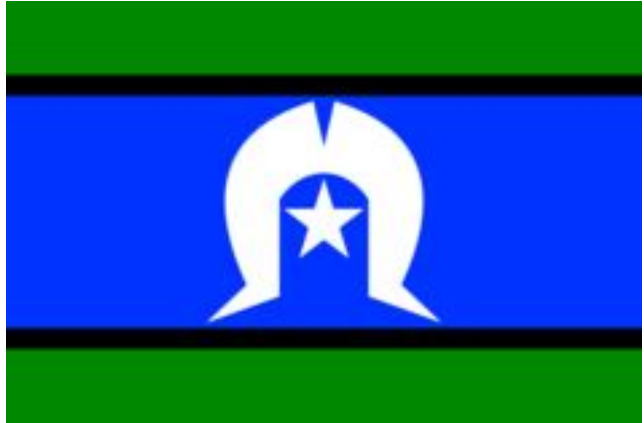
*Torres Strait flag. Credit: Bernark Namok.*

There are two distinct language groups:

- Meriam Mir, part of the Papuan language family, is spoken in the eastern islands

- Kala Lagaw Ya, part of the Australian language family, is spoken in the western, central, and northern islands

Islander culture stretches back nearly 8,000 years, when rising sea levels flooded the land bridge between Australia and Papua New Guinea at the end of the last Ice Age. Islanders in the northern and eastern island groups are primarily farmers. Those in the western groups rely heavily on hunting and fishing, and those in the central group rely mostly on trade with the other island groups.

This shows us that while Islanders share a common way of life, they are a diverse people spread out over a diverse geographic region with very long ancestral links to their country.

Torres Strait Islander culture is closely linked to the stars. They inform Islander laws, customs and practices that are recorded and handed down in the form of story, song, dance, ceremony and artefacts.

Islander astronomy also contains practical information about the natural world, which is essential for survival and cultural continuity. Islander culture is linked to Tagai – the creation deity that is represented by a constellation of stars that spans across the southern sky.

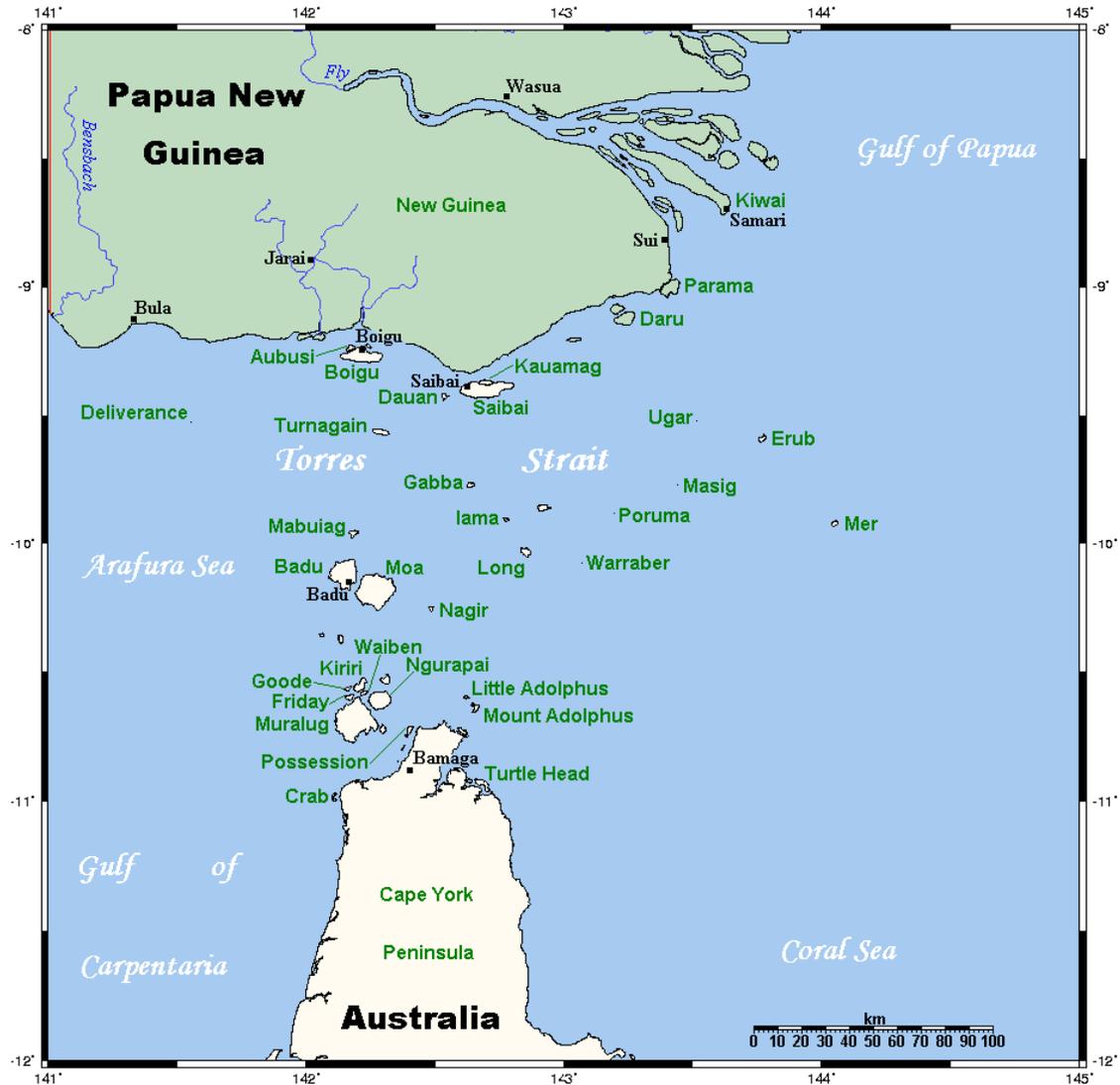

*Torres Strait and islands. Credit: Wikimedia Commons*

**Tagai's story**

Tagai was a great fisherman. One day he and his crew of 12 were fishing from their outrigger canoe. They were unable to catch any fish, so Tagai left the canoe and went onto the nearby reef to look for fish there.

As the day grew hotter and hotter, the waiting crew of Zugubals (beings who took on human form when they visited Earth) grew impatient and frustrated. Their thirst grew, but the only drinking water in the canoe belonged to Tagai. Their patience ran out and they drank Tagai's water.

When Tagai returned, he was furious that the Zugubals had consumed all of his water for the voyage. In his rage he killed all 12 of his crew. He returned them to the sky and placed them in two groups: six men in Usal (the Pleiades star cluster) and the other six Utimal (Orion). He told his crew to stay in the northern sky and to keep away from him.

Tagai can be seen in the southern skies, standing in a canoe in the Milky Way. His left hand is the Southern Cross holding a spear. His right hand is a group of stars in the constellation Corvus holding a fruit called Eugina. He is standing on his canoe, formed by the stars of Scorpius.

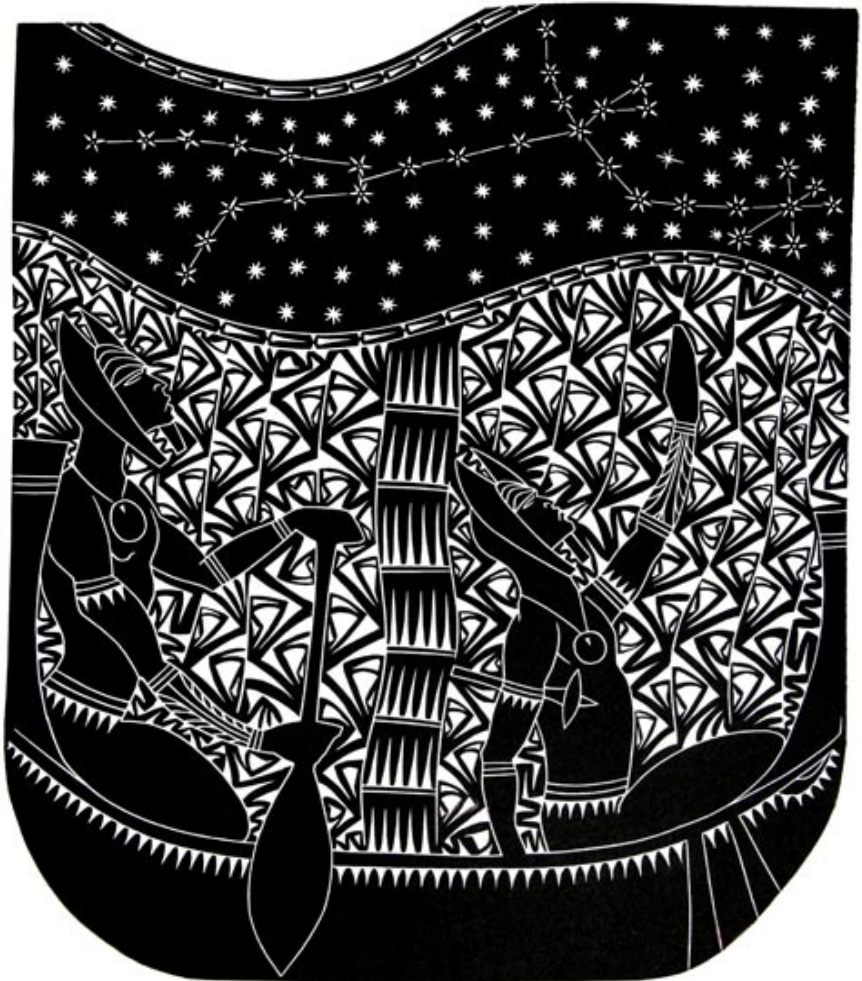

*Tagai. Credit: Glen Mackie*

**Tagai, today**

Islanders today still consider Tagai and astronomy to be an important aspect of daily life. Tagai is important for navigation, as the Southern Cross (his left hand) points in the direction of south.

The stars tell Islanders when to plant their gardens, when to hunt turtle and dugong, when the monsoon season arrives, when the winds change, and many other important aspects of daily life.

For example, when Tagai's left hand (the Southern Cross) dips into the sea, Islanders know the wet season (Kuki) is about to begin. The rising of Usal and Utimal (Pleiades and Orion) in mid-November tells Islanders that turtle and dugong are mating and that it's time to plant their gardens in anticipation of the coming Kuki season.

The shark constellation, Baidam, is made up of the stars in the Big Dipper, part of the constellation Ursa Major (the "Big Bear"). When these stars appear in the north over New Guinea, Islanders know the mating season of the shark is starting and that they should plant banana, sugar cane, and sweet potato. Lunar phases inform the Islanders of the best times to fish.

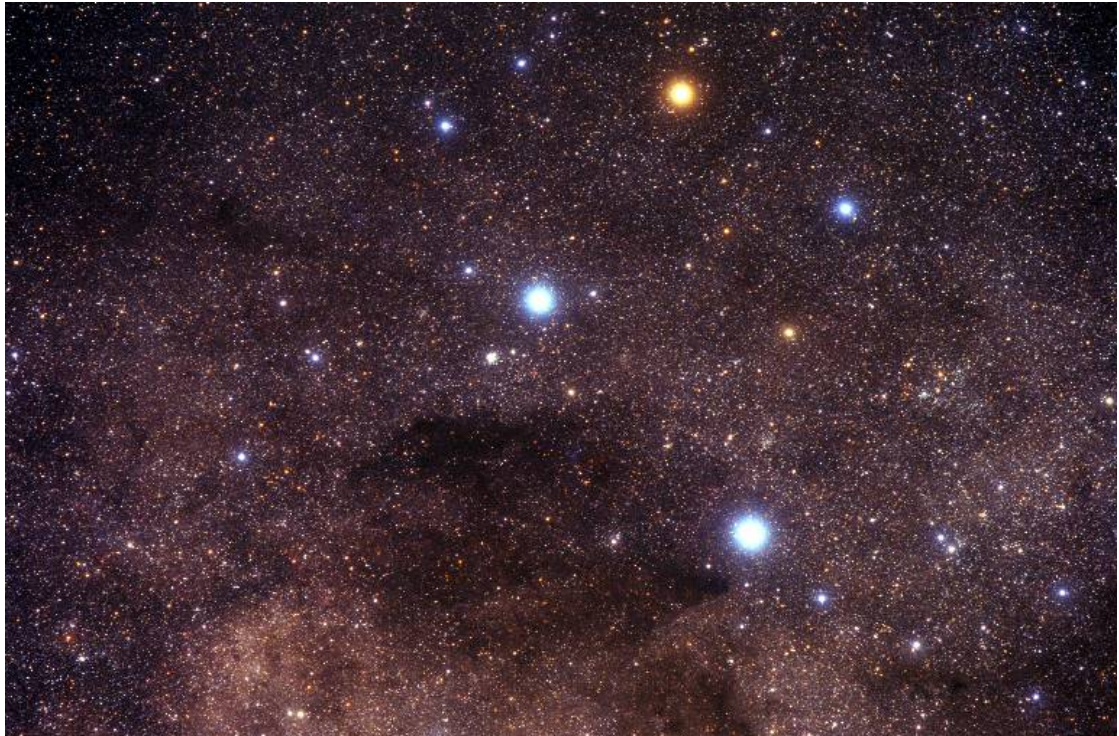
*Deep exposure of the Southern Cross – Tagai's left hand. Credit: Wikimedia Commons*

Islander astronomical knowledge is deep, and we are currently trying to better understand it for the benefit of future Islander generations.

*This article was originally published in The Conversation on 10 July 2013.*

## References

Sharp, N. 1993. *The Stars of Tagai.* Aboriginal Studies Press, Canberra.

Beckett, J. 1987. *The Torres Strait Islanders: custom and colonialism.* Cambridge University Press, Sydney